\documentclass[pre,aps,showpacs,twocolumn]{revtex4}
\usepackage{epsfig}

\newcommand{\be}{\begin{equation}}
\newcommand{\ee}{\end{equation}}
\newcommand{\bc}{\begin{center}}
\newcommand{\ec}{\end{center}}
\newcommand{\bi}{\begin{itemize}}
\newcommand{\ei}{\end{itemize}}
\newcommand{\ba}{\begin{eqnarray}}
\newcommand{\ea}{\end{eqnarray}}

\newcommand{\ignore}[1]{}
\newcommand{\mean}[1]{\left\langle #1 \right\rangle}

\begin{document}

\title{Voter model dynamics in complex networks: Role of dimensionality,
disorder and degree distribution.}

\author{Krzysztof Suchecki}
\altaffiliation[Present address: ]
{Center of Excellence for Complex Systems Research and Faculty of Physics,
Warsaw University of Technology,
 Koszykowa 75, PL-00-662, Warsaw, Poland}
\author{V\'{\i}ctor M. Egu\'{\i}luz}
\email{victor@imedea.uib.es}
\homepage{http://www.imedea.uib.es/physdept}
\author{Maxi San Miguel}
\email{maxi@imedea.uib.es}
\homepage{http://www.imedea.uib.es/physdept}
\affiliation{Instituto Mediterr\'aneo de Estudios Avanzados IMEDEA (CSIC-UIB),
E07122 Palma de Mallorca, Spain}

\date{\today}

\begin{abstract}

We analyze the ordering dynamics of the voter model in different classes of
complex networks. We observe that whether the voter dynamics orders the system
depends on the effective dimensionality of the interaction networks. We also
find that when there is no ordering in the system, the average survival time of
metastable states in finite networks decreases with network disorder and degree
heterogeneity. The existence of hubs in the network modifies the linear system
size scaling law of the survival time. The size of an ordered domain is
sensitive to the network disorder and the average connectivity, decreasing with
both; however it seems not to depend on network size and degree heterogeneity. 

\end{abstract}

\maketitle

\section{Introduction}

Equilibrium order-disorder phase transitions, as well as
nonequilibrium transitions and the kinetics of these transitions
\cite{Gunton83} have been widely study by spin Ising-type models
in different lattices \cite{Marro}. Given the recent widespread
interest on complex networks \cite{Albert02,Faloutsos99,Newman03,Eguiluz05} the effect of the
network topology on the ordering processes described by these
models has also been considered
\cite{Aleksiejuk02,Dorogovtsev02,Leone02,Bianconi02,Boyer}. In
particular models of opinion formation, or with similar social
motivations, have been discussed when interactions are defined
through a complex network
\cite{Eguiluz99,Sznaj,Klemm,Slanina,Klemm03a,DiscreteDeffaunt,Eguiluz05b}.

A paradigmatic and simple model where a systematic study of
network topology effects can be addressed is the voter model
\cite{Liggett85}, for which analytical and well established
results exist in regular lattices \cite{Krapivsky02,regularvoter}.
The dynamics of ordering processes for the voter model in regular
lattices \cite{Dornic01} is known to depend on dimensionality,
with metastable disordered states prevailing for $d>2$. In this
paper we address the general question of the role of network
topology in determining if the systems orders or not, and on the
dynamics of the ordering process. Specifically, analyzing the
voter model in several different networks, we consider the role of
the effective dimensionality of the network, of the degree
distribution and of the level of disorder present in the network.

The paper is organized as follows. In section 2 we shortly review
the basics as well as recent results on the voter model. Section 3
considers the voter model in scale free (SF) networks
\cite{Albert02} of different effective
dimensionality\cite{Eguiluz03}, showing that voter dynamics can
order the system in spite of a SF degree distribution. In section 4
we consider the role of network disorder by introducing a disorder
parameter that leads from a structured (effectively
one-dimensional) SF (SSF) network \cite{victornet1} to a random SF
(RSF) network through a small world \cite{Watts98} SF (SWSF)
network. The role of the degree distribution is discussed comparing
the results on the SSF, RSF and SWSF networks with networks with an
equivalent disorder but without a power law degree distribution.
Some general conclusions are given in Section 5.

\section{Voter Model}

The voter model \cite{Liggett85} is defined by a set of ``voters"
with two opinions or spins $\sigma_i= \pm 1$ located at the nodes
of a network. The elementary dynamical step consists in randomly
choosing one node (asynchronous update) and assigning to it the
opinion, or spin value, of one of its nearest neighbors, also
chosen at random. In a general network two spins are nearest
neighbors if they are connected by a direct link. Therefore, the
probability that a spin changes is given by
\be
P(\sigma_i \rightarrow -\sigma_i)=\frac{1}{2} \left(
1-\frac{\sigma_i}{k_i} \sum_{j\in {\cal V}_i} \sigma_j \right)~,
\ee
where $k_i$ is the degree of node $i$, that is the number of its
nearest neighbors, and ${\cal V}_i$ is the neighborhood of node $i$, that
is the set of nearest neighboring nodes of node $i$. In the
asynchronous update used here, one time step corresponds to
updating a number of nodes equal to the system size, so that each
node is, on the average, updated once. In our work we choose
initial random configurations with the same proportion of spins
$+1$ and $-1$.

The dynamical rule implemented here corresponds to a
\emph{node-update}. An alternative dynamics is given by a
\emph{link-update} rule in which the elementary dynamical step
consists in randomly choosing a pair of nearest neighbor spins,
{\it i.e.} a link, and randomly assigning to both nearest neighbor
spins the same value if they have different values, and leaving
them unchanged otherwise. These two updating rules are equivalent
in a regular lattice, but they are different in a complex network
in which different nodes have different number of nearest
neighbors \cite{Suchecki04}. In particular, both rules conserve
the ensemble average magnetization in a regular lattice, while in
a complex network this is only a conserved quantity for
link-update dynamics. Node-update dynamics conserves an average
magnetization weighted by the degree of the node \cite
{Suchecki04,Huberman}. We restrict ourselves in this paper to the
standard node-update for better comparison with the growing
literature on the voter model in complex networks
\cite{Castellano03,Vilone04,Redner04,Castellano05}.

The voter model dynamics has two absorbing states, corresponding
to situations in which all the spins have converged to the
$\sigma_i= 1$ or to the $\sigma_i= - 1$ states. The ordering
dynamics towards one of these attractors in a one-dimensional
lattice is equivalent to the one of the zero temperature kinetic
Ising model with Glauber dynamics. In more general situations, as
in regular lattice of higher dimension or in a complex network,
the ordering dynamics is still a zero-temperature dynamics driven
by interfacial noise, with no role played by surface tension. A
comparison of the voter model and the zero temperature Ising
Glauber dynamics in complex networks \cite{Boyer} has been
recently reported \cite{Castellano05}. A standard order parameter
to measure the ordering process in the voter model dynamics
\cite{Dornic01,Castellano03} is the average interface density
$\rho$, defined as the density of links connecting sites with
different spin value:
\be \rho = \left( \sum_{i=1}^N \sum_{j\in {\cal V}_i}
\frac{1-\sigma_i \sigma_j}{2} \right) / \sum_{i=1}^N k_i~.
\ee

\begin{figure}
\centerline{\epsfig{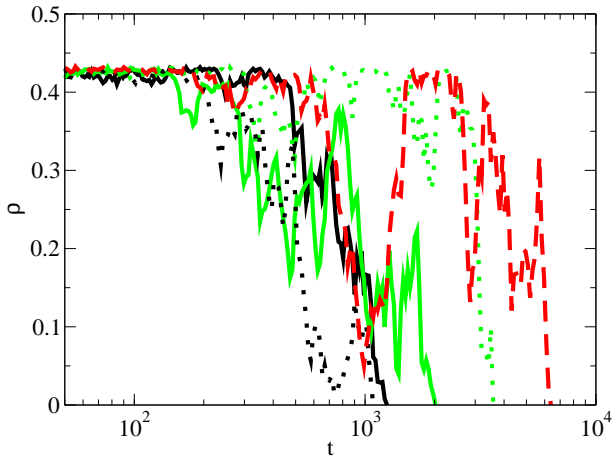}}
\caption{\label{ba_metastability}
Interface density evolution for individual realizations, for system
size $N=10000$ and average connectivity $\mean{k}=8$.}
\end{figure}

In a disordered configuration with randomly distributed spins
$\rho \simeq 1/2$, while when $\rho$ takes a small value it
indicates the presence of large spatial domains in which each spin
is surrounded by nearest neighbor spins with the same value. For a
completely ordered system, that is, for any of the two absorbing
states, $\rho=0$. Starting from a random initial condition, the
time evolution of $\rho$ describes the kinetics of the ordering
process. In regular lattices of dimensionality $d<2$ the system
orders. This means that, in the limit of large systems, there is a
coarsening process with unbounded growth of spatial domains of one
of the absorbing states. The asymptotic regime of approach to the
ordered state is characterized in $d=1$ by a power law $\langle
\rho \rangle \sim t^{-\frac{1}{2}}$, while for the critical
dimension $d=2$ a logarithmic decay is found $\langle \rho \rangle
\sim (\ln t)^{-1}$ \cite{Dornic01}. Here the average $\langle
\cdot \rangle$ is an ensemble average.

\begin{figure}
\centerline{\epsfig{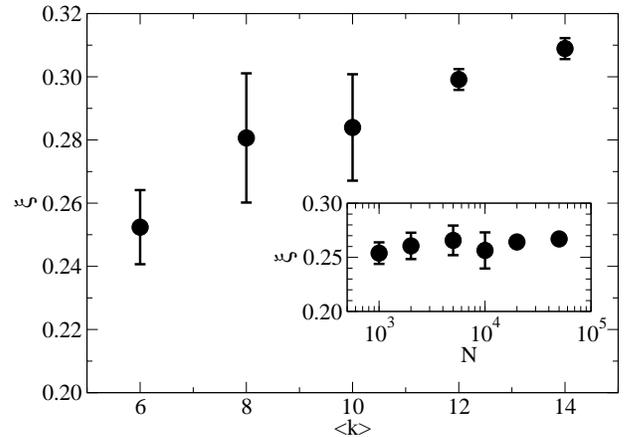}}
\caption{\label{ba_inset_plat}
Plateau height for SF networks with different average connectivity. Network size
$N=10000$. Inset: Plateau height dependence on network size. Average
connectivity $\mean{k}=6$. Data is averaged over $1000$ realizations.}
\end{figure}

In regular lattices with $d>2$ \cite{Krapivsky02}, as well as in
small world networks \cite{Castellano03}, it is known that the
voter dynamics does not order the system in the thermodynamic
limit of large systems. After an initial transient, the system
falls in these cases in a metastable partially ordered state where
coarsening processes have stopped: spatial domains of a given
attractor, on the average, do not grow. In the initial transient
of a given realization of the process, $\rho$ initially decreases,
indicating a partial ordering of the system. After this initial
transient $\rho$ fluctuates randomly around an average plateau
value $\xi$. This quantity gives a measure of the partial order of
the metastable state since $l=\xi^{-1}$ gives an estimate of the
average linear size of an ordered domain in that state. In a
finite system the metastable state has a finite lifetime: a finite
size fluctuation takes the system from the metastable state to one
of the two ordered absorbing states. In this process the
fluctuation orders the system and $\rho$ changes from its
metastable plateau value to $\rho=0$. Considering an ensemble of
realizations, the ordering of each of them typically happens
randomly with a constant rate. This is reflected in an exponential
decay of the ensemble average interface density
\be \mean{\rho} \propto e^{-\frac{t}{\tau}}~,
\label{exp}
\ee
where $\tau$ is the survival time of the partially ordered
metastable state. Note then that the average plateau value $\xi$
has to be calculated at each time, averaging only over the
realization of the ensemble that have not yet decayed to $\rho=0$.

The survival time $\tau$, for a regular lattice in $d=3$
\cite{Krapivsky02} and also for a small world network
\cite{Castellano03}, is known to scale linearly with the system
size $N$, $\tau \sim N$, so that the system does not order in the
thermodynamic limit. More recently the same scaling has been found
for random graphs \cite{Redner04,Castellano05} while a scaling
$\tau \sim N^{0.88}$ has been numerically found
\cite{Suchecki04,Castellano05} for the voter model in the scale
free Barabasi-Albert network \cite{Barabasi}. This scaling is
compatible with the analytical result $\tau \sim N/\ln N$ reported
in Ref.~\cite{Redner04}. Other analytical results for uncorrelated
networks with arbitrary power law degree distribution are also
reported in Ref.~\cite{Redner04}. We note that a conceptually
different, but related quantity, is the time that a finite system
takes to reach an absorbing state when coarsening processes are at
work. This time $\tau_1$ is known to scale as $\tau_1 \sim N^2$
for a regular $d=1$ lattice and $\tau_1 \sim N \ln N$ for a
regular $d=2$ lattice.

\begin{figure}
\centerline{\epsfig{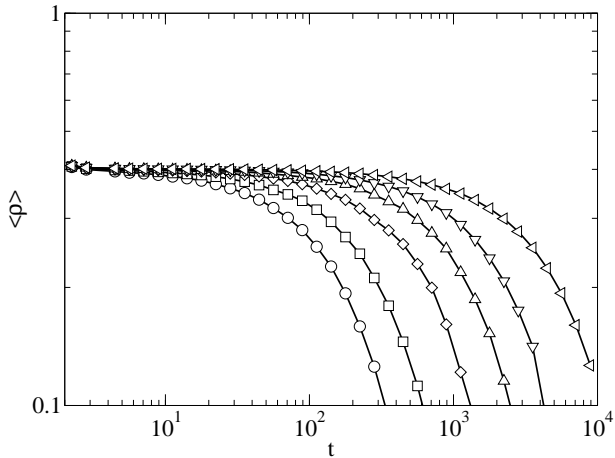}}
\caption{\label{ba_inter}
Interface density evolution in SF networks of different sizes (increasing from
left to right: $N=1000$, $2000$, $5000$, $10000$, $20000$, and $50000$). Data is averaged
over at least $1000$ realizations for $\mean{k}=6$.}
\end{figure}

In the next sections we discuss the time evolution of $\rho$ and
the characteristic properties of the plateau value $\xi$ and
survival time $\tau$ for asynchronous node-update voter dynamics
in a variety of different complex networks.

\section{Dimensionality and Ordering: Voter model in scale-free
networks}

One of simplest models that displays a scale-free degree
distributions is the  well known Barabasi-Albert network
\cite{Barabasi}. In this model, the degree distribution follows a
power law with an exponent $\gamma=3$, the path length grows
logarithmically with the system size \cite{Albert02} while the
clustering coefficient decreases with system size
\cite{victornet2}. It has been shown that critical phenomena on
this class of networks are well reproduced by mean field
calculations valid for random networks \cite{Goltsev03}. Thus we
will consider in the remainder the Barab\'asi-Albert networks as a
representative example of a random scale-free (RSF) network.
Results for the voter model in the BA network are shown in Figs.
\ref{ba_metastability}-\ref{ba_inter}. The qualitative behavior
that we observe is the same than the one described above for
regular lattices of $d>2$ or also observed in a small world network
\cite{Castellano03}: The system does not order but reaches a
metastable partially ordered state. The interface density $\rho$
for different individual realizations of the dynamics is shown in
Fig.~\ref{ba_metastability}. In this figure we see examples of how
finite size fluctuations take the system from the metastable state
with a finite plateau value of $\rho$ to the absorbing state with
$\rho=0$. The level of ordering in this finite lifetime metastable
state can be quantified by the plateau level $\xi$ shown in
Fig.~\ref{ba_inset_plat}. We find that the level of ordering
decreases significantly with the average connectivity of the
network, a result consistent with the idea that total ordering is
more easily achieved for effective lower dimensionality. On the
other hand the level of ordering is not seen to be sensitive to the
system size, for large enough sizes.

\begin{figure}
\centerline{\epsfig{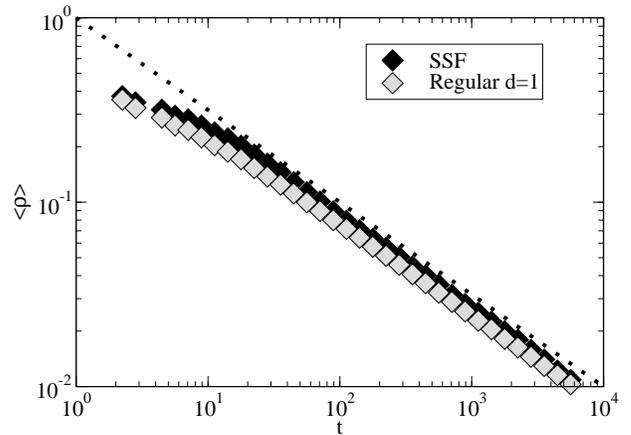}}
\caption{\label{1d_sf_nosf}
Interface density evolution in one dimensional systems with and without
scale-free topology. For reference, the dotted line is a power law with exponent
$-\frac{1}{2}$.Average over $1000$ realizations, $N=10000$ and $\mean{k}=8$.}
\end{figure}

The survival time $\tau$ can be calculated from the ensemble
average interface density $\mean{\rho}$ as indicated in
Eq.~(\ref{exp}). The time dependence of $\mean{\rho}$ for systems
of different size (Fig.~\ref{ba_inter}) shows an exponential
decrease for which the result mentioned above $\tau \sim N^{0.88}$
can be obtained \cite{Suchecki04}. We note that the value $\tau$
is found to be independent of the mean connectivity of the network
and that a linear scaling $\tau \sim N$ is obtained if a link
update dynamics is used \cite{Suchecki04}.

The fact that the presence of hubs in the BA network is not an efficient
mechanism to order the system might be counterintuitive, in the same way that
the presence of long range links in a small world network is also not efficient
to lead to an ordered state. However, in both cases the effective dimensionality
of the network is infinity and the result is in agreement with what is known for
regular lattices with $d>2$. A natural question is then the relevance of the
degree distribution versus the effective dimensionality in the ordering
dynamics. To address this question we have chosen to study the voter model
dynamics in the Structured Scale Free (SSF) network introduced in
Ref.~\cite{victornet1}. The SSF networks are a nonrandom network with a power
law degree distribution with exponent $\gamma=3$ but with an effective dimension
$d=1$ \cite{Eguiluz03}.

Our results for the time dependence of the average interface
density in the SSF network are shown in Fig.~\ref{1d_sf_nosf}. For
comparison the results for a regular $d=1$ network are also
included. For both networks we observe that the system orders with
the average interface density decreasing with a power law with
characteristic exponent $1/2$
\be
\mean{\rho} (t) \sim t^{-\frac{1}{2}}~.
\ee

The only noticeable difference is that the SSF network has a
larger number of interfaces at any moment, but the ordering
process follows the same power law. Additionally we find that for
a finite systems the time $\tau_1$ to reach the absorbing state
scales as $\tau_1\sim N^2$, as it also happens for the regular
$d=1$ network:

The network is completely ordered when the last interface
disappears. At this point, the density is simply $(N \mean{k})^{-1}$,
where $N \mean{k}$ is the total number of
links in the network. Since the interface density decreases
$\mean{\rho} \sim t^{-\frac{1}{2}}$, then the time to order
$\tau_1$ is given by
\be
(N \mean{k})^{-1}= \tau_1^{-1/2}~,
\ee
leading to $\tau_1 \sim N^2$.

\begin{figure}
\centerline{\epsfig{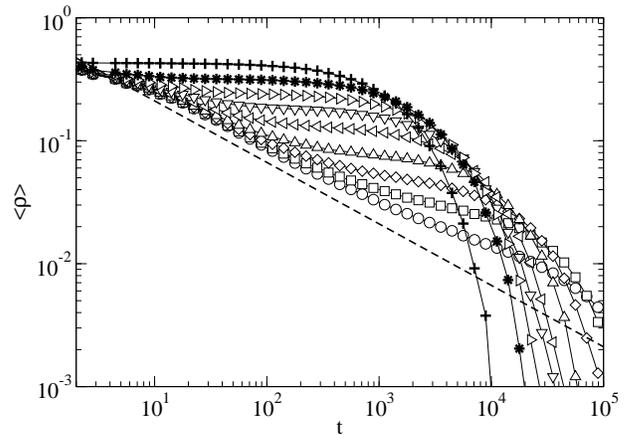}}
\caption{\label{1dsfsw_diffp}
Mean interface density evolution for SWSF networks with different disorder
parameter $p$ (increasing from circles to stars: $p=0.0004$, $0.001$,
$0.002$, $0.004$, $0.01$, $0.02$, $0.04$, and $0.1$), and comparison with the BA network (+). For reference, the dotted
line is a power law with exponent $-\frac{1}{2}$. Data averaged over $100$
realization for $p<0.01$ and over $1000$ realizations for $p\geq 0.01$;
$N=10000$ and $\mean{k}=8$.}
\end{figure}

Therefore we conclude that the effective dimensionality of the
network is the important ingredient in determining the ordering
process that results from a voter model dynamics, while the fact
that the system orders or falls in a metastable state is not
sensitive to the degree distribution.

\section{Role of network disorder and degree heterogeneity}

Once we have identified in the previous section the crucial role of
dimensionality we now address the role of network disorder and
degree heterogeneity in quantitative aspects of the voter model
dynamics. We do that by considering a collection of complex
networks in which the system falls into partially disordered
metastable states, except for the the regular one-dimensional lattice and SSF
networks in which the system shows genuine ordering dynamics:

\begin{enumerate}

\item Structured Scale Free (SSF) network as defined in the
previous section.

\item Small-World Scale-Free (SWSF) network. This is defined by rewiring with
probability $p$ the links of a SSF network. In order to conserve the degree
distribution of the unperturbed ($p=0$) networks, a randomly chosen link
connecting nodes $i,$ $j$ is permuted with that connecting nodes $k$, $l$)
\cite{Maslov03}. 

\item Random Scale Free (RSF) network: Defined as the limit $p=1$
of the SWSF network. The RSF network shares most important
characteristics with the BA network.

\begin{figure}
\centerline{\epsfig{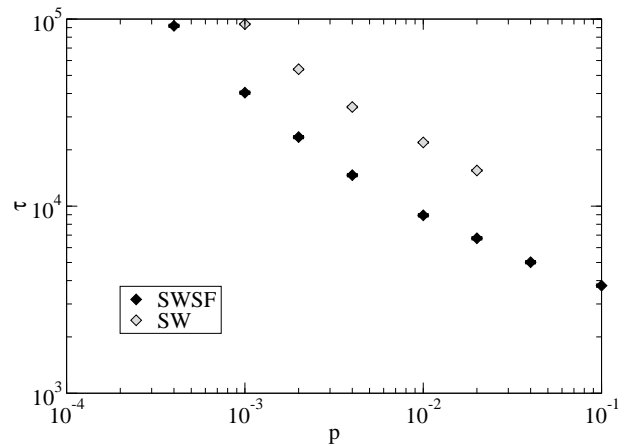}}
\caption{\label{1dsfsw_survp}
Survival times for a SWSF network with different disorder parameter $p$. For
comparison results for SW networks are also included. Average over $1000$
realizations, $N=10000$ and $\mean{k}=8$.}
\end{figure}

By changing the parameter $p$ from $p=0$ (SSF) to $p=1$ (RSF) we
can analyze how increasing levels of disorder affect the voter
model dynamics while keeping a scale free degree distribution. On
the other hand, the consequences of the degree heterogeneity
characteristic of SF networks can be analyzed comparing the voter
model dynamics on these networks with networks with the same level
of disorder and a non-SF degree distribution. These other networks
are constructed introducing the same disorder parameter $p$, but
starting from a regular $d=1$ network. Namely we consider:

\item Regular $d=1$ network that can be compared with a SSF
network.

\item Small World (SW) network defined introducing the rewiring
parameter $p$ in the regular network as in the prescription by
Watts and Strogatz \cite{Watts98}. The SW network can be compared
with the SWSF network.

\item Random network (RN) corresponding to the limit $p=1$ of the SW
network.

Likewise, one can consider a random network with an exponential
(EN) degree distribution. The EN network is constructed as in the
BA prescription but with random instead of preferential attachment
of the new nodes. These two random networks, RN and EN, can be
compared with the RSF network.

\end{enumerate}

\begin{figure}
\centerline{\epsfig{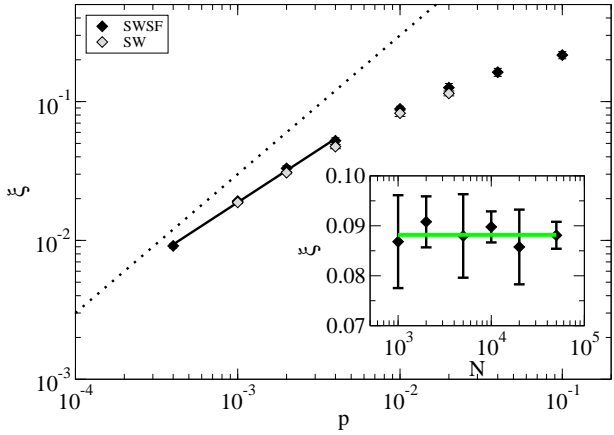}}
\caption{\label{1dsfsw_inset_plat}
Plateau heights for SWSF networks with different disorder parameter $p$. For
reference, dotted line is a power law $\xi \sim p$ while the solid line $\xi
\sim p^{0.76}$. For comparison results for SW networks
are also included. Average over $100$ realization for $p<0.01$ and over $1000$
realizations for $p\geq 0.01$, $N=10000$ and $\mean{k}=8$. Inset: Plateau
heights for SWSF networks of different system size $N$. Average over $1000$
realizations, with $p=0.01$ and $\mean{k}=8$.}
\end{figure}

\subsection{Role of disorder}

Figure~\ref{1dsfsw_diffp} shows the evolution of the mean interface
density for SWSF networks with different values of the disorder
parameter $p$. It shows how varying $p$ one smoothly interpolates
between the results for the SSF network and those for a RSF
network. In general, increasing network randomness by increasing
$p$ the system approaches the behavior in a BA network, making it
to to fall in a metastable state of higher disorder, but with
finite size fluctuations causing faster ordering. This trend is
quantitatively shown in Fig.~\ref{1dsfsw_survp} and
Fig.~\ref{1dsfsw_inset_plat} where the the survival time $\tau$ and
plateau level $\xi$ for SWSF networks are plotted as a function of
the disorder parameter $p$. We observe that $\tau$ and the size of
the ordered domains $l=\xi ^{-1}$ decrease with $p$ but without
following any clear power law. As a general conclusion, when
extrapolating to $p=1$, we find that $\tau_{SWSF}>\tau_{RSF}$ and
$l_{SWSF}>l_{RSF}$.

The role of increasing disorder in the network can also be
analyzed in networks without a scale free degree distribution by
considering SW networks with different values of the rewiring
parameter $p$. The survival time $\tau$ and plateau level $\xi$
for SW networks are also plotted in Fig.~\ref{1dsfsw_survp} and
Fig.~\ref{1dsfsw_inset_plat}. We observe that the effect of
disorder is qualitatively the same for SW than for SWSF networks
\cite{powerlawSW}. Extrapolating the results in
Fig.~\ref{1dsfsw_survp} and Fig.~\ref{1dsfsw_inset_plat} to $p=1$
where the SW network becomes a RN we find that
$\tau_{SW}>\tau_{RN}$ and $l_{SW}>l_{RN}$.

\begin{figure}
\centerline{\epsfig{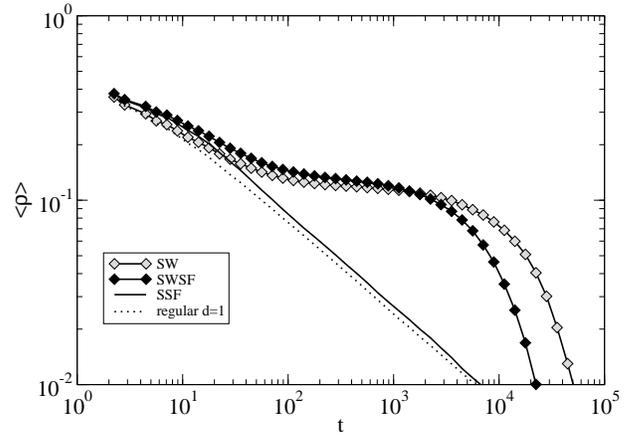}}
\caption{\label{1dsw_sf_nosf}
Interface density evolution for SFSW and SW networks with the same level of
disorder $p=0.01$. Average over $1000$ realizations, $N=10000$, and
$\mean{k}=8$.}
\end{figure}

\subsection{Role of degree distribution}

To address the question of the role of the degree distribution of
the network in the voter model dynamics we compare the evolution
in networks with a scale free degree distribution with the
evolution in equivalent networks but with a degree distribution
involving a single scale. A first comparison was already made
between the dynamics in a regular $d=1$ network and the SSF
network (Fig.~\ref{1d_sf_nosf}). This is included for reference in
Fig.~\ref{1dsw_sf_nosf} where we compare the evolution of the mean
interface density in a SWSF network with the evolution in a SW
network with the same level of disorder. We observe for SWSF a
similar plateau value (similar but slightly more disordered state)
at any time before the exponential decay of $\langle \rho \rangle$ which is
faster for the SWSF than for the SW networks. Finite size
fluctuations that order the system seem to be more efficient when
hubs are present, causing complete ordering more often, and
therefore a faster exponential decay of $\langle \rho \rangle$. These claims are
made quantitative in Fig.~\ref{1dsfsw_survp} and
Fig.~\ref{1dsfsw_inset_plat} where it is shown that
$\tau_{SW}>\tau_{SWSF}$ and $l_{SW} \simeq l_{SWSF}$. In addition,
extrapolating to the limit $p=1$ we have that
$\tau_{RN}>\tau_{RSF}$ and $l_{RN} \simeq l_{RSF}$

\begin{figure}
\centerline{\epsfig{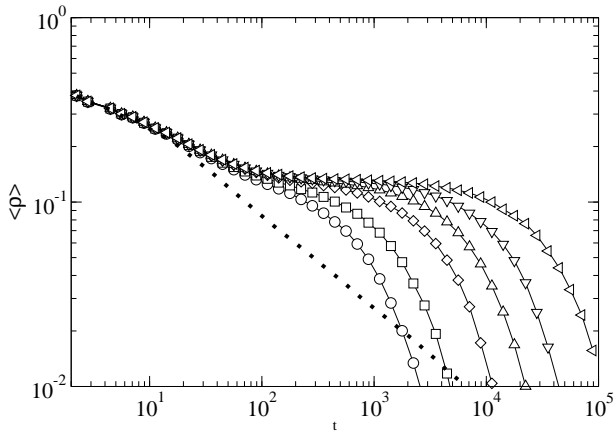}}
\caption{\label{1dsfsw_diffN}
Mean interface evolution for SWSF networks of different system size $N$
(increasing from left to right: $N=1000$, $2000$, $5000$, $10000$, $20000$,
$50000$). Average over $1000$ realizations, with $p=0.01$ and $\mean{k}=8$.}
\end{figure}

\begin{figure}
\centerline{\epsfig{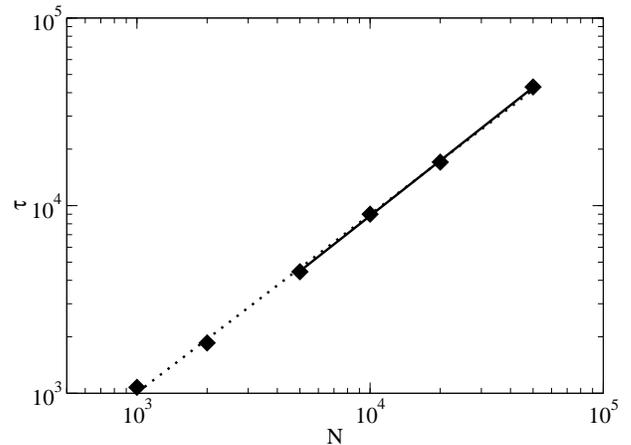}}
\caption{\label{1dsfsw_survN} Survival times for SWSF networks of
different system size $N$. Average over $1000$
realizations, with $p=0.01$ and $\mean{k}=8$.}
\end{figure}

It is also interesting to compare the dependence with system size
of the voter model dynamics in SW \cite{Castellano03} and SFSW
networks: The time dependence of the mean interface density for a
SWSF network with an intermediate fixed value of $p$ is shown in
(Fig.~\ref{1dsfsw_diffN}). The qualitative behavior is the same
than the one found for SW networks. However, the survival times
shown in Fig.~\ref{1dsfsw_survN} deviates consistently from the
linear power law $\tau \sim N$ found for SW networks
\cite{Castellano03}. This deviation might possibly have the same
origin than the deviation from the linear power law observed for
BA networks, that is the lack of conservation of magnetization in
the node-update dynamics of the voter model in a complex network
\cite{Suchecki04}. This non-conservation becomes much more
important in a SWSF network than in a SW network because of the
high degree heterogeneity. On the other hand we note that the
analytical results for survival times in Ref.~\cite{Redner04} apply
only to uncorrelated networks and therefore do not help us in
understanding our numerical result for SWSF networks. We also
mention that the plateau level $\xi$ for SWSF networks does not
show important dependence with system size (see inset of
Fig.~\ref{1dsfsw_inset_plat})

\begin{figure}
\centerline{\epsfig{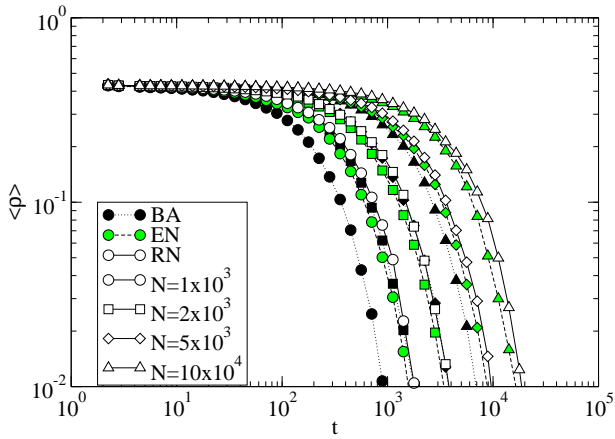}}
\caption{\label{exp_diffN}
Mean interface density evolution for RN and EN of different sizes $N$. BA
network of size $10000$ is also shown for comparison. Average over $1000$
realizations, and $\mean{k}=8$.}
\end{figure}

\begin{figure}[t]
\vskip 0.3cm
\centerline{\epsfig{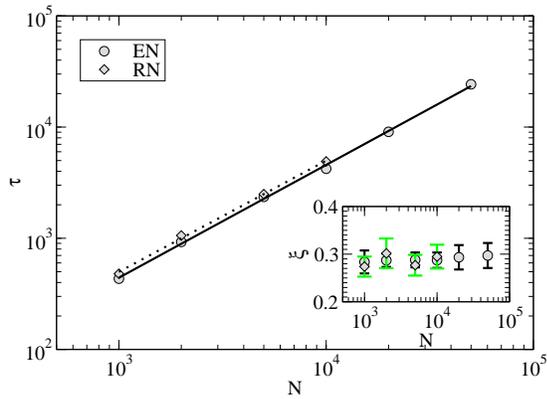}}
\caption{\label{exp_insetN} Survival times for RN and EN networks
of different sizes $N$. Average over $1000$
realizations, with $\mean{k}=8$. Inset: Plateau heights for the same
networks.}
\end{figure}

The role of degree heterogeneity can be further clarified
considering the limit of random networks $p=1$ where the SW
network becomes a RN and the SWSF network becomes a RSF network
essentially equivalent to the BA network. The evolution for the
mean interface density for different random networks is shown in
Fig.~\ref{exp_diffN}. We find again that when there are hubs (large
degree heterogeneity) there is a faster exponential decay of
$\langle \rho \rangle$, so that ordering is faster in BA networks
than in RN or EN, while the plateau level or level of order in
that state does not seem to be sensitive to the degree
distribution . This coincides with the extrapolation to $p=1$ of
the data in Fig.~\ref{1dsfsw_survp} and Fig.~\ref{1dsfsw_inset_plat}
which indicates that $\tau_{RN}>\tau_{RSF}$ and $l_{RN} \simeq
l_{RSF}$. Our results for the system size dependence of the
survival times and plateau levels for RN and EN networks are shown
in Fig.~\ref{exp_insetN}. The size of the ordered domains $l=\xi
^{-1}$ is again found not to be sensitive to system size. The
survival times for RN and EN networks follow a linear scaling
$\tau \sim N$ in agreement with the prediction in
Ref.~\cite{Redner04}. We recall that, as discussed earlier, in
random networks with scale free distribution such as the BA
network a different scaling is found ($\tau \sim N^{0.88}$)
\cite{Suchecki04,Castellano05} compatible with the prediction
$\tau \sim N/\ln N$ \cite{Redner04}.

\section{Conclusions}

We have analyzed how the ordering dynamics of the voter model is affected by the
topology of the network that defines the interaction among the nodes.
First we have shown that the voter model dynamics orders the system in a SSF
network \cite{victornet1}, which is a scale-free network with an effective
dimension $d=1$. This result, together with the known result that in regular
lattices the voter model orders in $d \le 2$, suggests that the effective
dimension of the underlying network is a relevant parameter to determine whether
the voter model orders, but not its degree distribution. The relevance of the
effective dimensionality of different scale-free networkshas also been observed
in other dynamical processes \cite{Eguiluz03,Warren02,Eguiluz02,Klemm03b}. In
the SSF network the density of interfaces in the voter model decreases as
$\langle \rho \rangle \sim t^{-1/2}$ in the same as in  the one-dimensional
regular lattices.

Second, we have introduced standard rewiring algorithms to study the effect of
network disorder. In general we find that network disorder decreases the
lifetime of metastable disordered states so that the survival time to reach an
ordered state in finite networks is smaller
$$
\tau_{SWSF} > \tau_{RSF},  \quad \tau_{SW} >\tau_{RS}~.
$$
Likewise, the average size of
ordered domains in these metastable states decreases with increasing disorder
$$
l_{SWSF} > l_{RSF},  \quad l_{SW} > l_{RS}~.
$$

Third, the degree heterogeneity also facilitates reaching an absorbing ordered
configuration in finite networks by decreasing the survival time: finite size
fluctuations ordering the system are more efficient when there are hubsin the
network, so that
$$
\tau_{SW}>\tau_{SWSF}, \quad \tau_{RN}>\tau_{RSF}~.
$$
The presence of hubs also invalidates the scaling law for the survival time
$\tau \sim N$ fouund in SW and RN. However we didn't appreciate differences in
the average size of ordered domains depending on degree heterogeneity
$$
l_{SW} \simeq l_{SWSF}, \quad l_{RN} \simeq l_{RSF}~.
$$

In summary, we find for the different classes of networks considered in this
work that
\ba
\tau_{SW}>&\tau_{RN}&>\tau_{RSF}      \nonumber       \\
l_{SW} \simeq l_{SWSF}>&l_{RN}& \simeq l_{RSF}~. 	\nonumber
\ea
In general our results illustrate how different features (dimensionality, order,
degree heterogeneity) of complex networks modify key aspects of a simple
stochastic dynamics.

\acknowledgments We acknowledge financial support from MEC (Spain)
through projects CONOCE2 (FIS2004-00953) and FIS2004-05073-C04-03.
KS thanks Prof. Janusz Holyst for very helpful comments.



\begin{thebibliography}{2}
\bibitem{powerlawSW} The power law dependence of $\xi$ on $p$
reported previously in Ref.~\cite{Vilone04} for SW networks is
included for reference in Fig.~\ref{1dsfsw_inset_plat}, but seems
to be valid only for very small values of $p$.
\end{thebibliography}
\end{document}